\documentclass[aps,prl,twocolumn,amsfonts,amssymb,amsmath,showpacs,
floatfix,nofootinbib,groupedaddress,superscriptaddress,citesort]{revtex4}
\usepackage{mathrsfs}
\usepackage{amsfonts}
\usepackage{amstext}
\usepackage{amsmath}
\usepackage{amssymb}
\usepackage{times}
\usepackage[dvips]{graphicx}
\def\qed{\leavevmode\unskip\penalty9999 \hbox{}\nobreak\hfill
     \quad\hbox{\leavevmode  \hbox to.77778em{%
              \hfil\vrule   \vbox to.675em%
               {\hrule width.6em\vfil\hrule}\vrule\hfil}}
     \par\vskip3pt}

\begin{document}
\title{Quantum advantage by weak measurements}

\author{Lin-Ping Huai}
\affiliation{Theoretical Computational Institute, School of Physics and Information Technology, Shannxi Normal University, Xi'an 710062, China}
\affiliation{Institute of Physics, Chinese Academy of Sciences, Beijing 100190, China}
 \author{Bo Li}
  \affiliation{Institute of Physics, Chinese Academy of Sciences, Beijing 100190, China}

\author{Shi-Xian Qu}
\email{sxqu@snnu.edu.cn}
\affiliation{Theoretical Computational Institute, School of Physics and Information Technology, Shannxi Normal University, Xi'an 710062, China}
\author{Heng Fan}
\email{hfan@iphy.ac.cn}
\affiliation{Institute of Physics, Chinese Academy of Sciences, Beijing 100190, China}

\begin{abstract}
Weak measurements may result in extra quantity of quantumness of correlations compared with standard projective measurement
on a bipartite quantum state. We show that the quantumness of correlations by weak measurements can be consumed for
information encoding which is only accessible by coherent quantum interactions.
Then it can be considered as a resource for quantum information processing and can quantify this quantum advantage.
We conclude that weak measurements can create more valuable quantum correlation.
\end{abstract}
\pacs{}
\maketitle

\section{Introduction}
Quantum information processing offers protocols and algorithms which may surpass their classical counterparts \cite{M.A.Nielsen}.
We generally consider that those quantum advantages are due to unique features of quantum world such as the
superposition and the quantumness of correlations like entanglement and quantum discord. Indeed, entanglement and
quantum discord are shown to be resources for some quantum information tasks. However, different criteria
are used to distinguish quantum from classical worlds and various measures are applied in quantifying
the quantum correlations.
Entanglement is a well studied quantum correlation and it plays a clear role in
Bell inequalities, teleportation and superdense coding ect. The history of entanglement study can trace
back to Einstein, Podolsky, Rosen who considered this spooky action at a distance in quantum mechanics \cite{Einstein}.

 However, entanglement seems not account for all quantum correlation in bipartite system since
 the absence of entanglement does not eliminate all signatures of quantum behaviour.
Henderson and Vedral \cite{Henderson}, Ollivier and Zurek \cite{Ollivier} proposed a quantity of  nonclassical correlation, which is named as quantum discord,
to measure the discrepancy of total and classical correlation. It is thus a measure of quantumness of correlation.
Quantum discord is shown to be present in a mixed state quantum computation algorithm, the deterministic quantum computation with one qubit (DQC1), which
offers the quantum advantage but with negligible entanglement.
It is then considered that quantum discord plays a key role in this algorithm.
Alternatively, some other proposals are also discussed DQC1 \cite{YuCS}.
In the past years, great progress have been made in understanding entanglement and various measures of quantum correlations
\cite{Brasssard,D.P.DiVincenzo,H.J.Bernstein,P.L.Knight,Plenio,Ollivier,Henderson,C.H.Bennett et al,Oppenheim,Horodecki,Popescu}.

Recently, it is found that discord consumption can be related with the quantum advantage for encoding information \cite{Mile Gu}.
This is a clear evidence that quantum discord can be considered as a resource for quantum information processing like the role
of entanglement in teleportation. In this protocol,
quantum advantage is upper bounded by the consumption of discord during the encoding information and is lower bounded by the difference of discord consumption and classical correlation. Thus, quantum discord consumed during encoding can quantify the advantage about coherent interactions.
This operational significance of discord has enlightened our knowledge of quantum correlation other than entanglement.

On the other hand, quantum discord is quantum measurement dependent.
Its quantity depends on the positive operator valued
measurement (POVM) or projection measurement \cite{M.A.Nielsen}. We know that quantum measurement is one of the
cornerstones of quantum mechanics. Instead of the strong measurement such as the projection measurement which will
cause completely collapsing of the wavefunction of a quantum state, a measurement scheme called ``weak measurement'', which only induces
partial collapsing of the quantum state, is indispensable \cite{aav,Korotkiv2006,Sun2009}.
As we known, weak measurement plays a key role in the fundamentals of quantum mechanics and
has practical applications \cite{kwiat,lundeen,SJWu}.
Now, it seems natural to consider a weak measurement scheme for quantum discord \cite{Uttam Singh}. In particular, it is very recently found that weak measurement
can recover the quantum discord and preserve the fidelity of quantum teleportation for some decoherence channels \cite{LiYanLing1,LiYanLing2,Pramanik}.
It is understandable that the accessible information by weak measurement will be smaller than that of a projection measurement which is
strong, so the residue correlation will become larger. The result is that we may obtain extra quantumness of
correlation by weak measurement. While the standard quantum discord is shown to be a valuable resource,
then the question is whether this new defined quantumness of correlation which is larger is also useful.
In this paper, we will provide a positive answer to this question! We can prove that
for two-qubit pure state and Bell-diagonal state, this super discord induced by weak measurement can indeed be consumed during the encoding information and therefore can be  regarded also as a precious resource in this quantum protocol.
Our finding reveals that more resources of quantum correlation might be explored by weak measurement.
This highlights the fundamental role of weak measurements in studying quantum correlations.

This paper is organized as follows.  In Section \uppercase\expandafter{\romannumeral2}, we first give a brief review of  the definition of quantum advantage,  the formulism of  the projective measurement and weak measurement. In Section \uppercase\expandafter{\romannumeral3},
we prove that super discord by weak measurement can achieve the best quantum advantage,
for which the quantum correlation is completely consumed, in the case of  general pure entangled states and the Bell-diagonal states.
Section \uppercase\expandafter{\romannumeral4} is the summary and discussion.

\section{Quantum correlation with different measurements  }\label{surface}
For a bipartite quantum state $\varrho_{ab}$ shared by Alice and Bob, in general, the total correlation between
the two subsystems is measured by mutual information,
\begin{eqnarray*}
I(\varrho_{ab})=S(\varrho_{a})+S(\varrho_{b})-S(\varrho_{ab}),
\end{eqnarray*}
where $S(\varrho_{x})=-{\rm Tr}(\varrho_{x}\log\varrho_{x})$ is the von Neumann entropy,
$\varrho_{x}$ is the density matrix of system $x$. Here, $\rho_{a}$ and $\rho_{b}$ are the reduced matrices of the state $\varrho_{ab}$,
$ \varrho_{a}={\rm Tr}_{b}\varrho_{ab}$, $\varrho_{b}={\rm Tr}_{a}\varrho_{ab}$.
Classical correlation accessible by measurement is defined as
\begin{eqnarray}
J(\varrho_{a,b})=S(\varrho_{a})-\mathrm{min}_{\{\Pi_{i}^{b}\}}\sum_{i}p_{i}S(\varrho_{a|\Pi_{i}^{b}}),
\label{classi}
\end{eqnarray}
where $\{\Pi_{i}^{b}\}$  is the set of all possible measurements that performed locally only on subsystem $b$ on Bob's side.
So the quantum discord\cite{Einstein} is defined as the difference between those two correlations
\begin{eqnarray}
\delta(a|b)=I(\varrho_{ab})-J(\varrho_{a,b}).
\label{discord}
\end{eqnarray}

For normal quantum discord, the measurement acting on one side refers to the POVM or projective measurement. A POVM with $d$ ($d$ is the dimension of system) outcomes is a $d$-tuple of operators$ (\Pi_{0}^{b},\Pi_{1}^{b},...,\Pi_{d-1}^{b})$, and $\Pi_{i}^{b}\geq{0}, \sum_{i}\Pi_{i}^{b}=\mathbb{I}$.
The conditional state $\varrho_{{a}|\Pi_{i}^{b}}={\rm Tr}_{b}(\Pi_{i}^{b}\varrho_{ab})/p_{i}$, $p_{i}={\rm Tr}_{ab}(\Pi_{i}^{b}\varrho_{ab})$ is the probability of the outcome $i$.
For two-qubit system, we know that the best measurement that Bob can perform to get information about Alice's system $a$ is a projective measurement onto the state of system $b$. Suppose $\{\Pi_{i}^{b}\}$ are the elements of projective measurement, we can assume $\Pi_{i}=|i\rangle\langle{i}|,i=0,1$, and $\sum_{i}\Pi_{i}=\sum_{i}\Pi_{i}^{\dagger}\Pi_{i}=\mathbb{I}$.

Very recently, Gu \emph{et al}. \cite{Mile Gu} consider the operational significance of discord consumption during encoding information within state $\varrho_{ab}$, where the discord is obtained by the POVM. In this paper, we mainly follow this scheme but apply it to the more general case.
Next, we present the details of the scheme.
Alice encodes an arbitrary feasible $K$ with probability $p_{k}$ onto her subsystem $a$ by application of unitary operator $U_{k}$.
After encoding, the state becomes  $\widetilde{\varrho}_{ab}=\sum_{k}p_{k}U_{k}\varrho_{ab}U_{k}^{\dagger}$ and is given to Bob.
Bob uses some decoding protocols to get the best possible estimate of the encoded data $K$ from $\widetilde{\varrho}_{ab}$.
The Holevo accessible information can used for this scheme.

Before proceed, we consider that Bob has no access to $b$ but can access $a$,
the amount of information accessible to Bob is given by, $I_{0}=S(\widetilde{\varrho}_{a})-S(\varrho_{a})$,
here $\widetilde{\varrho}_{a}={\rm Tr}_{b}(\widetilde{\varrho}_{ab})=\sum_{k}p_{k}U_{k}\varrho_{a}U_{k}^{\dagger}$.
This case can be considered as that Bob cannot access any of the correlations between the two subsystems.
Next, the scheme is divided into two different cases.
First, Bob can perform any measurements on the whole system $\widetilde{\varrho}_{ab}$ to retrieve the encoded random
variable $K$. For this case, $\{I_{q}-I_{0}\}$ stands for the information that Bob can obtain, where $I_{q}=S(\widetilde{\varrho}_{ab})-S(\varrho_{ab})$.
Second, Bob is allowed to implement only the local measurement on $a$ or $b$.
If Bob firstly measures on system $b$ then $a$, he can obtain the maximal information defined as, $\overleftarrow{I}_{c}=\mathrm{sup}_{{\{\Pi_{i}^{b}\}}}(\sum_{i}p_{i}S(\widetilde{\varrho}_{a|\Pi_{i}^{b}})-\sum_{i}p_{i}S(\varrho_{a|\Pi_{i}^{b}}))$, here $S(\widetilde{\varrho}_{a|\{\Pi_{i}^{b}\}})$ is the  quantum conditional entropy.
Alternatively, Bob first measures on $a$ then $b$, $\overrightarrow{I}_{c}$ is the information that Bob can obtain and $\overrightarrow{I}_{c}\leq{I_{0}+S(\varrho_{b})-\mathrm{min}_{\{\Pi_{i}^{a}\}}\sum_{i}S(\varrho_{b|\Pi_{i}^{a}})}$,
$\{\Pi_{i}^{a}\}$ means measurement on system $a$. We take the maximum value between $\overleftarrow{I}_{c}$ and $\overrightarrow{I}_{c}$ as the optimal information $I_{c}$.

We can find that $\triangle{I}=I_{q}-I_{c}$ is the difference of obtained information
between coherent measurements and local measurements.
It can be considered as the extra quantum advantage which is induced by only coherent interactions in the whole process.
It is proven that the following inequality  must hold for any $\varrho_{ab}$ under any feasible measurements,
\begin{eqnarray}
\triangle{\delta{(a|b)}}-\widetilde{J}(\varrho_{a,b})\leq{\triangle{I}}\leq{\triangle{\delta(a|b)}}.
\end{eqnarray}
We know that $\triangle{\delta{(a|b)}}$ and $\widetilde{J}(\varrho_{ab})$ represent the amount of discord consumed during encoding and classical correlation after encoding respectively. If there is no discord between Alice and Bob, then the quantum advantage cannot exist.
If the encoding is the maximal encoding, we have $I(\sum_{k}p_{k}U_{k}\varrho_{ab}U_{k}^{\dagger})=0$,
$\varrho_{ab}=\widetilde{\varrho}_{a}\otimes{\widetilde{\varrho}_{b}}, \widetilde{\varrho}_{a}$ is a maximally mixed state regardless of $\widetilde{\varrho}_{b}$.
So the classical correlation of the state after encoding $\widetilde{J}$ is zero.
For the maximal encoding, the quantum advantage is exactly equal to the discord consumed during encoding.
For this case, we consider the quantum advantage as the best quantum advantage.
When Alice and Bob have no discord at the beginning, the quantum advantage is meaningless.
In that way, coherent interactions will be no use for Bob to retrieve the encoded information.

Aharonov-Albert-Vaidman \cite{Y.Aharonov} came up with the weak measurement formalism,
where the system interacts weakly with the measured tool.
The system is disturbed weakly and the coherence may not be destroyed completely.
This is in sharp contrast with the projective measurement which is strong and results in complete decoherence of
the system. However, any projective measurement can be decomposed into a sequence of weak measurements,
which change the quantum state gradually and in the end result in the same outcomes.
Further more, any measurement can be generated by weak measurements, so weak measurements are universal \cite{O.Oreshkov}.
Weak measurement plays both a fundamental role in quantum theory and can have physical
applications \cite{N.W.M.Ritchie,G.J.Pryde,O.Hosten,J.S.Lundeen,N.S.Williams}.
Quantum discord can also be defined for weak measurement and it is shown to be larger
than the standard quantum discord based on projective measurements \cite{Uttam Singh}.

For weak measurement, a measurement with n-tuples of outcomes can be simplified to having two outcomes, and
the weak measurement operators are given as \cite{O.Oreshkov}
  \begin{eqnarray*}
\widehat{P}(x)=\sqrt{X_{-}}\widehat{P}_{1}+\sqrt{X_{+}}\widehat{P}_{2},
\end{eqnarray*}
 \begin{eqnarray}
\widehat{P}(-x)=\sqrt{X_{+}}\widehat{P}_{1}+\sqrt{X_{-}}\widehat{P}_{2},
\label{p}
\end{eqnarray}
where, $x\in{R}$, representing the strength of the measurement process, and $X_{-}=\frac{1-\tanh{x}}{2}, X_{+}=\frac{1+\tanh{x}}{2}$, if $x=\varepsilon$, when $|\varepsilon|\ll{1}$, $\widehat{P}(\pm{x})$ is weak measurement, $\widehat{P}_{1}$ and $\widehat{P}_{2}$ are orthogonal projectors with $\widehat{P}_{1}+\widehat{P}_{2}=\widehat{I}, \widehat{P}^{2}(x)+\widehat{P}^{2}(-x)=\widehat{I}$, $\widehat{I}$ is the identity.
For the superposition state $|\psi\rangle=\cos{\frac{\theta}{2}}|0\rangle+e^{i\varphi}\sin{\frac{\theta}{2}}|1\rangle$,
we can let $\widehat{P}_{1}=|\psi\rangle\langle{\psi}|$, $\widehat{P}_{2}=|\psi^{\bot}\rangle\langle{\psi^{\bot}}|$,
where $|\psi^{\bot}\rangle$ are orthogonal to $|\psi\rangle $.  Explicitly, the operators $\widehat{P}_{1}$ and $\widehat{P}_{2}$ are
\begin{displaymath}
   \begin{split}
  \widehat{P}_{1}&=\cos^{2}{\frac{\theta}{2}}|0\rangle\langle{0}|+e^{-i\varphi}\sin{\frac{\theta}{2}}\cos{\frac{\theta}{2}}|0\rangle\langle{1}|\\
              &+e^{i\varphi}\sin{\frac{\theta}{2}}\cos{\frac{\theta}{2}}|1\rangle\langle{0}|+\sin^{2}{\frac{\theta}{2}}|1\rangle\langle{1}|,\\
   \end{split}
   \end{displaymath}
 \begin{displaymath}
   \begin{split}
  \widehat{P}_{2}&=\cos^{2}{\frac{\theta}{2}}|1\rangle\langle{1}|+e^{-i\varphi}\sin{\frac{\theta}{2}}\cos{\frac{\theta}{2}}|0\rangle\langle{1}|\\
              &-e^{i\varphi}\sin{\frac{\theta}{2}}\cos{\frac{\theta}{2}}|1\rangle\langle{0}|+\sin^{2}{\frac{\theta}{2}}|0\rangle\langle{0}|,\\
   \end{split}
   \end{displaymath}
Under the condition of the maximal encoding, we will show that for different measurements,
the best quantum advantages are different.

\section{The best quantum advantage in specific families of states}\label{dynamics}
We introduce the generalized Pauli matrices $U_{m,n}=X^{m}Z^{n}, m,n=0,...,d-1$ in $d$ dimension, where
$X|j\rangle=|j+1\mathrm{mod}d\rangle, Z|j\rangle=w^{j}|j\rangle, w=e^{2\pi{i}/d}$, here $\{|j\rangle\}_{0}^{d-1}$
is an orthonormal basis.
These operators constitute a basis of unitary operators, see for example \cite{Heng Fan}.
When $d=2$, we recover the standard Pauli matrices,
\begin{displaymath}
X=
\left(\begin{array}{ccc}
0&1\\
1&0
\end{array}\right),
Y=iXZ=
\left(\begin{array}{ccc}
0&-i\\
i&0
\end{array}\right),Z=
\left(\begin{array}{ccc}
1&0\\
0&-1
\end{array}\right).
\end{displaymath}
For a general pure entangled state of bipartite system
\begin{eqnarray}
|\phi\rangle_{ab}=\sum_{i}\sqrt{\lambda_{i}}|i_{a}\rangle|i_{b}\rangle,
\label{ab}
\end{eqnarray}
where $\lambda_{i}$ are Schmidt coefficients and satisfy $\sum_{i}\lambda_{i}=1$.

\emph{The two-qubit pure state.}---
In the following, we will restrict our attentions on two-dimensional case.
The two-qubit pure state studied is $|\phi\rangle=\sqrt{\lambda_{0}}|00\rangle+\sqrt{\lambda_{1}}|11\rangle$.
We first consider the projective measurement acting on the system, and
follow the scheme presented in Ref.\cite{Mile Gu}.
The quantum discord of $|\phi\rangle$ is given as,
\begin{eqnarray}
\delta(a|b)=S(\varrho_{a})=-\lambda_{0}\log\lambda_{0}-\lambda_{1}\log\lambda_{1}.
\end{eqnarray}
We consider the encoding as applying Pauli matrices and identity with equal probability,
so the state after encoding becomes as,
\begin{eqnarray}
\begin{split}
\widetilde{\varrho}_{ab}&=\frac{1}{2}\sum_{m,n=0}^{1}(X^{m}Z^{n}\otimes{I})\varrho_{ab}(X^{m}Z^{n}\otimes{I})^{\dagger}\\
&=\frac{I_{a}}{2}\otimes{(\lambda_{0}|0\rangle_{b}\langle{0}|+\lambda_{1}|1\rangle_{b}\langle{1}|)}.\\
\end{split}
\label{abe}
\end{eqnarray}
 From Eq.(\ref{abe}), we can find, $\widetilde{J}=\widetilde{\delta}=0$, where $\widetilde{J}$ and $\widetilde{\delta}$
are the classical correlation and the discord of the state after encoding, respectively.
 The amount of discord consumed during encoding is $\triangle{\delta}=\delta-\widetilde{\delta}=-\lambda_{0}\log\lambda_{0}-\lambda_{1}\log\lambda_{1}$.
  Next we calculate quantum advantage,
  here the measurement acts on $b$ of the state $\widetilde{\varrho}_{ab}$, the minimum of condition entropy is given as $\mathrm{min}_{\{\Pi_{i}^{b}\}}\sum_{i}p_{i}S(\widetilde{\varrho}_{a|\Pi_{i}^{b}})=1$.
From the encoding scheme, we obtain
\begin{displaymath}
\begin{split}
I_{q}&=1-\lambda_{0}\log\lambda_{0}-\lambda_{1}\log\lambda_{1},\\
I_{c}&=1.\\
\end{split}
\end{displaymath}
Thus the quantum advantage is obtained as,
 \begin{eqnarray}
\triangle{I}_{p}=-\lambda_{0}\log\lambda_{0}-\lambda_{1}\log\lambda_{1},
\label{p}
\end{eqnarray}
We can find that the consumption of quantum discord $\triangle{\delta}$
is equal to quantum advantage.
Thus it is the best quantum advantage which corresponds to the maximum encoding.

As an example, for maximally entangled pure state $|\phi\rangle =(|00\rangle +|11\rangle )/\sqrt {2}$,
we have $\widetilde{\varrho}_{ab}=\frac{I_{a}}{2}\otimes{\frac{I_{b}}{2}}, I_{q}=2, I_{c}=1$,
the best quantum advantage and the consumption of quantum discord is $1$.

Now let us turn to the weak measurement. The measurement operator acting on the system is given as
\begin{eqnarray}
I_{a}\otimes{\widehat{P}_b(x)}&=&A(x,\theta)(|00\rangle\langle{00}|+|10\rangle\langle{10}|)+B(x,\theta,\varphi)
                                                                                                    \nonumber\\
              &&(|00\rangle\langle{01}|+|10\rangle\langle{11}|)+C(x,\theta,\varphi)(|01\rangle\langle{00}|
                                                                                                 \nonumber\\
              &&+|11\rangle\langle{10}|)+D(x,\theta)(|01\rangle\langle{01}|+|11\rangle\langle{11}|),
                                                                                          \nonumber\\
\end{eqnarray}
where $A(x,\theta)=\sqrt{X_{-}}\cos^{2}\frac{\theta}{2}+\sqrt{X_{+}}\sin^{2}\frac{\theta}{2}$, $B(x,\theta)=(\sqrt{X_{-}}-\sqrt{X_{+}})e^{-i\varphi}\sin{\frac{\theta}{2}}\cos{\frac{\theta}{2}}$, $C(x,\theta)=B^{*}(x,\theta)$, $D(x,\theta)=\sqrt{X_{-}}\sin^{2}\frac{\theta}{2}+\sqrt{X_{+}}\cos^{2}\frac{\theta}{2}$.
Similarly we can obtain the other coefficients when $x$ is negative, the probability is, $p(\pm{x})={\rm Tr}(\widehat{P}_{b}(\pm{x})^{\dagger}\widehat{P}_{b}(\pm{x})\varrho)=\frac{1}{2}(1-(\lambda_{0}-\lambda_{1})\tanh{(\pm{x})}\cos{\theta})$. After measurement the eigenvalues are
    \begin{eqnarray*}
k_{1}(\pm{x})=\frac{1}{2}[1+{\sqrt{1-\frac{\lambda_{0}\lambda_{1}}{p(\pm{x})^{2}\cosh^{2}x}}}],\\ k_{2}(\pm{x})=\frac{1}{2}[1-{\sqrt{1-\frac{\lambda_{0}\lambda_{1}}{p(\pm{x})^{2}\cosh^{2}x}}}].
\end{eqnarray*}
Thus the quantum conditional entropy for weak measurement is given as
 \begin{eqnarray*}
 S_{w}(\widetilde{\varrho}_{a|\{M_{b}(x)\}})&=&p(x)S(\widetilde{\varrho}_{a|M_{b}(x)})+p(-x)S(\widetilde{\varrho}_{a|M_{b}(-x)})\\
                                              &=&1.
\end{eqnarray*}
According to the encoding scheme presented in Eq.(\ref{abe}), Bob implements now weak measurement,
the maximal information obtained by Bob by different coherent or local measurements, $I_{q}$ and $I_{c}^{x}$, are given as
\begin{eqnarray*}
I_{q}&=&-2\lambda_{0}\log\lambda_{0}-2\lambda_{1}\log\lambda_{1},\\
I_{c}^{x}&=&-\lambda_{0}\log\lambda_{0}-\lambda_{1}\log\lambda_{1}+p(x)[k_{1}(x)\log{k_{1}(x)}\\
     &&+k_{2}(x)\log{k_{2}(x)}]+p(-x)[k_{1}(-x)\log{k_{1}(-x)}\\
     &&+k_{2}(-x)\log{k_{2}(-x)}].
\end{eqnarray*}
The best quantum advantage which is the difference of the above two quantities can be found as,
 \begin{eqnarray}
\triangle{I}_{w}&=&-\lambda_{0}\log\lambda_{0}-\lambda_{1}\log\lambda_{1}-p(x)[k_{1}(x)\log{k_{1}(x)}
                                                                                                  \nonumber\\
&&+k_{2}(x)\log{k_{2}(x)}]-p(-x)[k_{1}(-x)\log{k_{1}(-x)}
                                                \nonumber\\
&&+k_{2}(-x)\log{k_{2}(-x)}].
\end{eqnarray}
This quantity depends on the initial pure entangled state and the strength of the weak measurement $x$.
When the state is a maximally entangled state, $I_{q}=2$, $I_{c}^{x}=1+X_{-}\log{X_{-}}+X_{+}\log{X_{+}}$.
 The best quantum advantage is
 \begin{eqnarray}
\triangle{I}_{w}&=2-X_{-}\log{(2X_{-})}-X_{+}\log{(2X_{+})}.
\label{x}
\end{eqnarray}

\begin{figure}
  \includegraphics[width=2.7in]{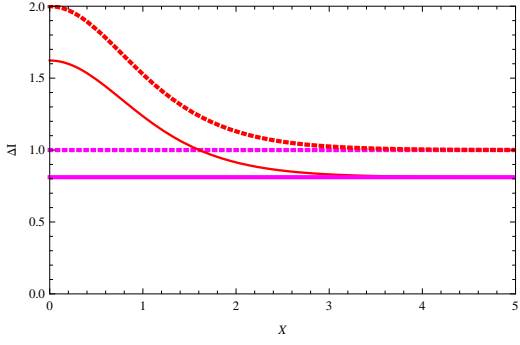}\\
  \caption{Colour online}Quantum advantage as a function of the measurement strength $x$ for maximally entangled state(dotted line) and
  the pure state($\lambda_{0}=\frac{\sqrt{2}}{2},\lambda_{1}=1-\frac{\sqrt{2}}{2}$(solid line). The straight lines represent projective measurement, and
  the bent lines represent weak measurement.
    \label{bentcure}
\end{figure}
In Fig.\ref{bentcure}, we plot different cases of quantum advantage in terms of the measurement strength $x$.
One can see that quantum advantage
revealed by weak measurement is always larger than that when only projective measurement is performed.
Also, quantum advantage is gradually diminishing
with the increasing of measurement strength $x$.
When $x$ is larger than approximately $2.7$,
the difference between $\triangle{I}_{w}$ and $\triangle{I}_{p}$ is negligible,
which means the coherent interaction does not provide much help to Bob.
One may notice that the maximally entangled state has much more best quantum advantage than the other pure states for both weak measurement and normal projective measurement.
So we may conclude that weak measurement
can reveal much more quantum advantage than normal projective measurement for general bipartite pure entangled state.

\emph{The two-qubit Bell diagonal state.}---
For the two-qubit Bell diagonal states \cite{D.Bruss}
\begin{eqnarray}
\varrho_{ab}=\frac{1}{4}(I+\sum_{j=1}^{3}c_{j}\sigma_{j}\otimes{\sigma_{j}}),
\label{x}
\end{eqnarray}
where $c_{j}\in\mathbb{R}$, when $|c_{1}|=|c_{2}|=|c_{3}|=c$, it is a Werner state,
in case $|c_{1}|=|c_{2}|=|c_{3}|=1$, it is one Bell state. Consider the optimumal projective measurement, the conditional entropy is \cite{S.Luo}
 \begin{eqnarray*}
S(\varrho_{a}|\{\Pi_{i}^{b}\})=\frac{1-c}{2}\log{(1-c)}+\frac{1+c}{2}\log{(1+c)},
\end{eqnarray*}
where $c=\mathrm{max}{\{|c_{1}|,|c_{2}|,|c_{3}|\}}$.
By maximal encoding, the state becomes, $\widetilde{\varrho}_{ab}=\frac{I_{a}}{2}\otimes{\frac{I_{b}}{2}}$.
By methods similar to the previous, we have $I_{q}$ and $I_{c}$ as follows,
\begin{eqnarray*}
I_{q}&=&\frac{1}{4}[(1-c_{1}-c_{2}-c_{3})\log{(1-c_{1}-c_{2}-c_{3})}+(1\\
&&+c_{1}+c_{2}-c_{3})\log{(1+c_{1}+c_{2}-c_{3})}+(1+c_{1}-\\
&&c_{2}+c_{3})\log{(1}{+c_{1}-c_{2}+c_{3})}+(1-c_{1}+c_{2}+c_{3})\\
&&\log{(1}{-c_{1}+c_{2}+c_{3})}],\\
I_{c}&=&1+\frac{1-c}{2}\log{\frac{1-c}{2}}+\frac{1+c}{2}\log{\frac{1+c}{2}}.\\
\end{eqnarray*}
The difference is the best quantum advantage, and it is written as,
 \begin{eqnarray}
\triangle{I}_{p}&=&\frac{1}{4}[(1-c_{1}-c_{2}-c_{3})\log{(1-c_{1}-c_{2}-c_{3})}{}
                                                        \nonumber\\
&&+(1+c_{1}+c_{2}-c_{3})\log{(1+c_{1}+c_{2}-c_{3})}+(1{}
                                                \nonumber\\
&&+c_{1}-c_{2}+c_{3})\log{(1+c_{1}-c_{2}+c_{3})}+(1-c_{1}{}
                                                 \nonumber\\
&&+c_{2}+c_{3})\log{(1}{-c_{1}+c_{2}+c_{3})}]-\frac{1-c}{2}\log{(1-c)}{}
                                                           \nonumber\\
&&-\frac{1+c}{2}\log{(1+c)}.{}
                   \nonumber\\
\end{eqnarray}

For weak measurements, the probability is $p(x)=\frac{1}{2}$,
\begin{eqnarray*}
I_{q}&=&\frac{1}{4}[(1-c_{1}-c_{2}-c_{3})\log{(1-c_{1}-c_{2}-c_{3})}\\
&&+(1+c_{1}+c_{2}-c_{3})\log{(1+c_{1}+c_{2}-c_{3})}+(1\\
&&+c_{1}-c_{2}+c_{3})\log{(1+c_{1}-c_{2}+c_{3})}+(1-c_{1} \\
&&+c_{2}+c_{3})\log{(1}{-c_{1}+c_{2}+c_{3})}],\\
I_{c}^{x}&=&1+\frac{1}{2}[\lambda_{1}(x)\log{\lambda_{1}(x)}+\lambda_{2}(x)\log{\lambda_{2}(x)}\\
&&+\lambda_{1}(-x)\log{\lambda_{1}(-x)}+\lambda_{2}(-x)\log{\lambda_{2}(-x)}],\\
\end{eqnarray*}
where the eigenvalues are $\lambda_{1}(\pm{x})=\frac{1+\sum_{j}c_{j}n_{j}\tanh(\pm{x})}{2}$, $\lambda_{2}(\pm{x})=\frac{1-\sum_{j}c_{j}n_{j}\tanh(\pm{x})}{2}$,
$n_{1}=\sin{\theta}\cos{\psi}$, $n_{2}=\sin{\theta}\sin{\psi}$, $n_{3}=\cos{\theta}$.

The best quantum advantage is
 \begin{eqnarray}
\triangle{I}_{w}&=&\frac{1}{4}[(1-c_{1}-c_{2}-c_{3})\log{(1-c_{1}-c_{2}-c_{3})}{}
                                                                   \nonumber\\
&&+(1+c_{1}+c_{2}-c_{3})\log{(1+c_{1}+c_{2}-c_{3})}{}
                                           \nonumber\\
&&+(1+c_{1}-c_{2}+c_{3})\log{(1+c_{1}-c_{2}+c_{3})}{}
                                            \nonumber\\
&&+(1-c_{1}+c_{2}+c_{3})\log{(1}{-c_{1}+c_{2}+c_{3})}]{}
                                              \nonumber\\
&&-1-\frac{1}{2}[\lambda_{1}(x)\log{\lambda_{1}(x)}+\lambda_{2}(x)\log{\lambda_{2}(x)}{}
                                                                         \nonumber\\
&&+\lambda_{1}(-x)\log{\lambda_{1}(-x)}+\lambda_{2}(-x)\log{\lambda_{2}(-x)}].{}
                                                                    \nonumber\\
\end{eqnarray}
\begin{figure}
  \includegraphics[width=2.7in]{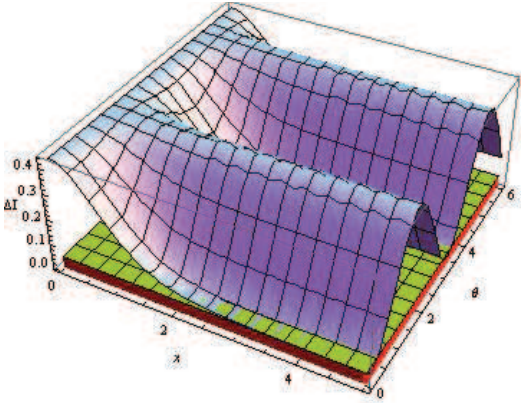}\\
  \caption{Color online}The best quantum advantage is a function of $x$ and $\theta$ for Bell-diagonal state with $c_{1}=0.15, c_{2}=0.03, c_{3}=0.7, \varphi=0$. The
  purple surface represent only using weak measurement, the green plane at $0.045$ represent quantum advantage using projective measurement.
\label{223}
\end{figure}
In Fig.\ref{223}, the quantum advantage induced by weak measurement is greater than that normal projective measurement.
When $\theta=n\pi(n=0,\pm{1},...)$, $x$ is larger than approximately $3$, the quantum advantage with weak measurement is close to that with normal projective measurement. Thus we summarize that for Bell-diagonal states, weak measurement reveal much more quantum advantage than the normal projective measurement. It is worth mentioning that for weak measurement,  $\theta$ possesses a periodic behavior (such as $\theta=0, \pm{\pi},...,n\pi$).

For Werner state, i.e., $|c_{1}|=|c_{2}|=|c_{3}|=c$. It is a mixture of the fully mixed state with probability $1-p$ and a singlet state $|\psi\rangle$ with probability $p$ \cite{D.Bruss},
\begin{eqnarray}
\varrho_{ab}=c|\psi\rangle\langle\psi|+\frac{I}{4}(1-c),
\label{ab}
\end{eqnarray}
here $|\psi\rangle=\frac{|01\rangle-|10\rangle}{\sqrt{2}}$. Depending on the encoding protocol mentioned above, we obtain the best quantum advantage with the normal projective measurement and weak measurement respectively,
 \begin{eqnarray}
\triangle{I}_{p}&=&\frac{1+3c}{4}\log{(1+3c)}+\frac{1-c}{4}\log{(1-c)}{}
                                                       \nonumber\\
&&-\frac{1+c}{2}\log{(1+c)},{}
                  \nonumber\\
\end{eqnarray}
 \begin{eqnarray}
\triangle{I}_{w}&=&\frac{3(1-c)}{4}\log{(1-c)}+\frac{1+3c}{4}\log{(1+3c)}-{}
                                                                        \nonumber\\
&&{}\frac{1-c\tanh{x}}{2}\log{(\frac{1-c\tanh{x}}{2})}-\frac{1+c\tanh{x}}{2}{}
                                                                         \nonumber\\
&&\log{(\frac{1+c\tanh{x}}{2})}-1.{}
                             \nonumber\\
\end{eqnarray}
\begin{figure}
  \includegraphics[width=2.7in]{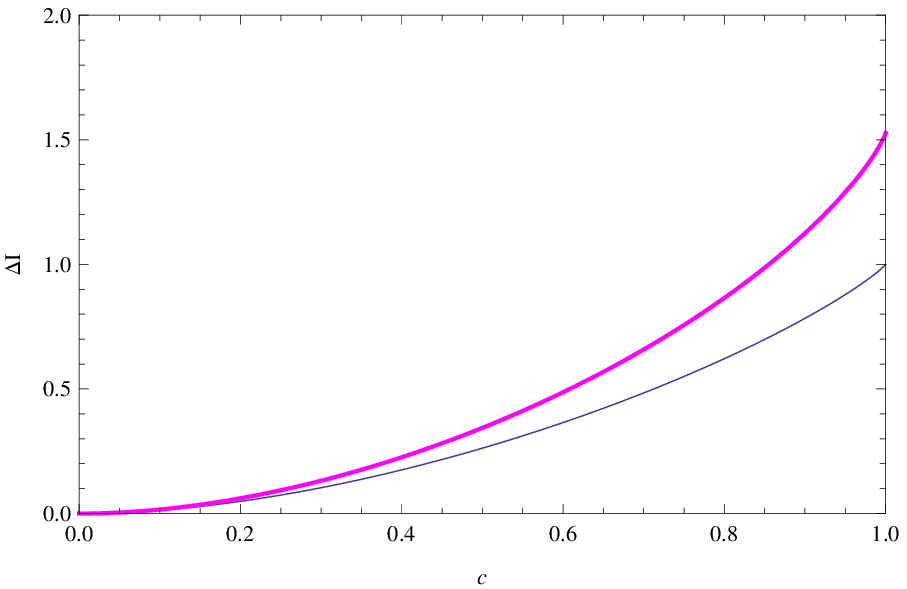}\\
  \caption{} The best quantum advantage for Werner state as function of $c$ when we fix the strength of the
  measurement $x=0.7$. The purple curve represent using weak measurement, while
  the gray curve means using projective measurement.
  \label{aaa}
  \end{figure}
\begin{figure}
  \includegraphics[width=2.7in]{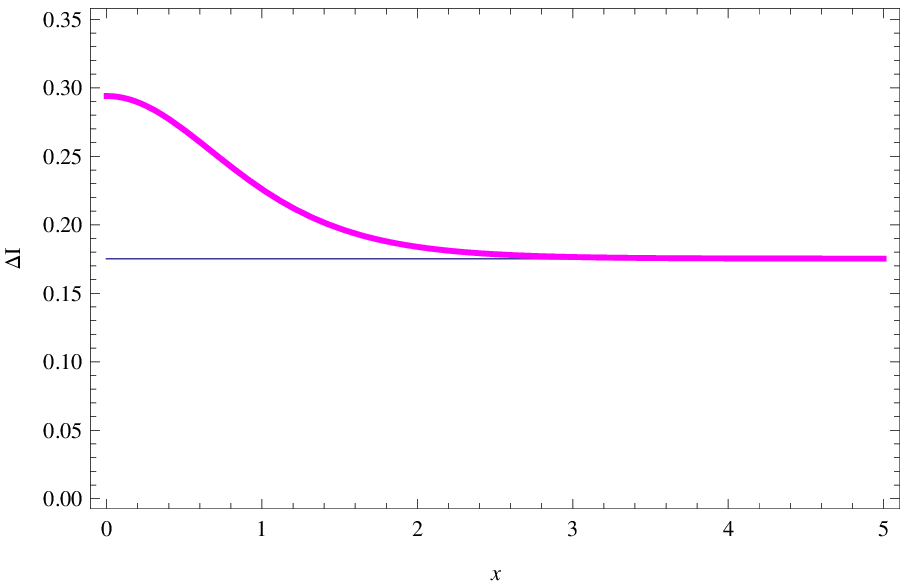}\\
  \caption{}The best quantum advantage for Werner state as a function of the measurement strength $x$ while the state
  is fixed to case $c=0.4$. The purple curve represents the quantum advantage with weak measurement, and the gray line means with normal projective measurement.
  \label{bbb}
\end{figure}
In Fig.\ref{aaa}, we plot quantum advantage for Werner state, the best quantum advantage by using weak measurements with fixed strength $x=0.7$
is greater than the one with projective measurements for the whole class of states, since the entanglement and quantum discord is monotonously increasing along with the increase of parameter $c$. One can also say that quantum advantage is a monotone function verses entanglement and quantum discord for Werner state.
Meanwhile, in Fig.\ref{bbb}, the state parameter $c=0.4$, when with weak measurement, quantum advantage is gradually diminishing along with the measurement strength $x$, when $x$ is smaller, quantum advantage is always larger than that when only projective measurement is performed.
But if $x$ is larger than around $2.5$, $\triangle{I}_{w}$ approach to $\triangle{I}_{p}$,
which means the coherent interactions almost have no help to Bob. These further illustrate that weak measurement
can reveal much more quantum advantage than normal projective measurement for Werner state.

\section{summary and discussions}\label{discuss}
In summary, weak measurement is of fundamental interest in quantum science.
It is also very useful to recover the quantum discord and preserve the
fidelity of teleportation, as shown in Refs.\cite{LiYanLing1,LiYanLing2,Pramanik}.
In this article, instead of protecting quantum correlation resources like
discord and entanglement, we show that the
additional quantumness of correlations induced by weak measurement
is not only a quantity which has the meaning of quantification of quantum correlation, but also it
may be the valuable resource in quantum information processing. One can see that
all of the quantum correlation based on weak measurement can be consumed for encoding information,
it may demonstrate the quantum advantage in comparing coherent measurements with local measurements.
Our result provides a new application of weak measurement in quantum information processing.

\bigskip
\noindent {\bf Acknowledgments}:
Linping Huai would like to thank  Li-Xing Jia and  Li-Ming Zhao for helpful discussions on this topic. This work is supported by NSFC (11175248,11305105), ``973'' program (2010CB922904).


\begin{thebibliography}{18}
\bibitem{M.A.Nielsen} M. A. Nielsen and I. L. Chuang, Quantum Computation and Quantum Information(Cambridge University Press, Cambridge, England, 2002).
\bibitem{Einstein} A. Einstein, B. Podolsky, and N.Rosen, Phys. Rev. {\bf 47}, 777 (1935).
\bibitem{Henderson} L. Henderson, and V. Vedral, J. Phys. A \textbf{34}, 6899 (2001).
\bibitem{Ollivier} H. Ollivier and W. H. Zurek, Phys. Rev. Lett. {\bf 88}, 017901 (2001).
\bibitem{YuCS}C. S. Yu, X. X. Yi, H. S. Song, and H. Fan, Phys. Rev. A {\bf 87}, 022322 (2013).
\bibitem{Brasssard} C. H. Bennett, G. Brasssard, C. Crepeau, R. Jozsa, A. Peres and W. K. Wootters, Phys. Rev. Lett. {\bf 70}, 1895 (1993).
\bibitem{D.P.DiVincenzo} C. H. Bennett, D. P. DiVincenzo, J. A. Smolin and W. K. Wootters, Phys. Rev. Lett. {\bf 54}, 3824 (1996).
\bibitem{H.J.Bernstein} C. H. Bennett, H. J. Bernstein, S. Popescu and B. Schumacher, Phys. Rev. A {\bf 78}, 2275 (1996).
\bibitem{P.L.Knight} V. Vedral, M. B. Plenio, M. A. Rippin and P. L. Knight, Phys. Rev. Lett. {\bf 78}, 2275 (1997).
\bibitem{Plenio} V. Vedral and M. B. Plenio, Phys. Rev. A {\bf 57}, 1619 (1998).
\bibitem{C.H.Bennett et al} C. H. Bennett et al., Phys. Rev. A {\bf 59}, 1070 (1999).
\bibitem{Oppenheim} J. Oppenheim, M. Horodecki, P. Horodecki, and R. Horodecki, Phys. Rev. Lett. {\bf 89}, 180402 (2002).
\bibitem{Horodecki} M. Horodecki et al., Phys. Rev. A {\bf 71}, 062307 (2005).
\bibitem{Popescu} B. Groisman, S. Popescu, and A.Winter, Phys. Rev. A {\bf 72}, 032317 (2005).
\bibitem{Mile Gu} M. Gu, Helen M. Chrzanowski, Syed M. Assad, Thomas Symul, Kavan Modi, Timothy C. Ralph, Vlatko Vedral and Ping Koy Lam, Nature. Phys. {\bf 10}.1038 (2012).
\bibitem{aav} Y. Aharonov, D. Z. Albert, and L. Vaidman, Phys. Rev. Lett. {\bf 60}, 1351 (1988).
\bibitem{Korotkiv2006} A. N. Korotkov and A. N. Jordan, \prl {\bf97}, 166805 (2006).
\bibitem{Sun2009} Q. Sun, M. Al-Amri, and M. S. Zubairy, \pra {\bf80}, 033838 (2009).
\bibitem{kwiat}O. Hosten and P. Kwiat, Science {\bf 319}, 787 (2008).
\bibitem{lundeen}J. S. Lundeen \emph {et al.}, Nature {\bf 474}, 188 (2011).
\bibitem{SJWu}S. J. Wu, Sci. Rep. {\bf 3}, 1193 (2013).
\bibitem{Uttam Singh}U. Singh and A. Kumar Pati, Annals of
Phys. {\bf 343}, 141 (2014).




\bibitem{LiYanLing1}Y. L. Li and X. Xiao, Quant. Inf. Process {\bf 12}, 3067 (2013).
\bibitem{LiYanLing2}X. Xiao and Y. L. Li, Eur. Phys. J. D {\bf 67}, 204 (2013).

\bibitem{Pramanik}T. Pramanik and A. S. Majumdar, Phys. Lett. A {\bf 377}, 3209 (2013).



\bibitem{Y.Aharonov} Y. Aharonov, D. Z. Albert, and L. Vaidman, Phys. Rev. Lett. {\bf 60}, 1351 (1988).
\bibitem{O.Oreshkov} O. Oreshkov and T. A. Brun, Phys. Rev. Lett. {\bf 95}, 110409 (2005).
\bibitem{N.W.M.Ritchie} N. W. M. Ritchie, J. G. Story, and R. G. Hulet, Phys. Rev. Lett. {\bf 66}, 1107 (1991).
\bibitem{G.J.Pryde} G. J. Pryde et al., Phys. Rev. Lett. {\bf 94}, 220405 (2005).
\bibitem{O.Hosten} O. Hosten and P. Kwiat, Science {\bf 319}, 787 (2008)
\bibitem{J.S.Lundeen} J. S. Lundeen and A. M. Steinberg, Phys. Rev. Lett. {\bf 102}, 020404 (2009).
\bibitem{N.S.Williams} N. S. Williams and A. N. Jordan, Phys. Rev. Lett. {\bf 100}, 026804 (2008).
\bibitem{Heng Fan} H. Fan, Phys. Rev. Lett. {\bf 92}, 177905 (2004).
\bibitem{D.Bruss} D. Bruss, J. Math. Phy. {\bf 43}, 4237 (2002).
\bibitem{S.Luo} S. Luo, Phys. Rev. A {\bf 77}, 042303 (2008).
\end{thebibliography}
\end{document}